\documentclass[twocolumn,aps,amssymb,prl]{revtex4}
\usepackage{graphicx}
\usepackage{epsfig,amsmath}
\usepackage{color}

\begin{document}

\title{Gate-controlled superconductivity in diffusive multiwalled carbon nanotube}

\author{T.~Tsuneta$^{1}$, L. Lechner$^{1,2}$, and P.~J.~Hakonen$^1$}
\affiliation{$^1$Low~Temperature~Laboratory,~Helsinki~University~of~Technology,~Finland
\\$^2$University of Regensburg, Regensburg, Germany}

\date{\today} 

\begin{abstract}
We have investigated electrical transport in a diffusive multiwalled
carbon nanotube contacted using superconducting leads made of Al/Ti
sandwich structure. We find proximity-induced superconductivity with
measured critical currents up to $I_{cm} = 1.3$ nA, tunable by gate
voltage down to 10 pA. The supercurrent branch displays a finite
zero bias resistance which varies as $R_0 \propto I_{cm}^{-\alpha}$
with $\alpha =0.74$. Using IV-characteristics of junctions with
phase diffusion, a good agreement is obtained with Josephson
coupling energy in the long, diffusive junction model of A.D Zaikin
and G.F. Zharkov (Sov. J. Low Temp. Phys. \textbf{7}, 184 (1981)).

\end{abstract}
\pacs{PACS numbers: 67.57.Fg, 47.32.-y} \bigskip

\maketitle

 Superconductivity in carbon nanotubes is an intriguing
subject. To understand it one has to consider many facets of modern
physics, including Luttinger liquid behavior owing to strong
electron-electron interactions in one dimension as well as Kondo
physics due to odd, unpaired electronic spin \cite{review}.
Intrinsic superconductivity has been observed in ropes of single
walled carbon nanotubes (SWNT) \cite{KasumovScience}, while
proximity induced superconductivity has been investigated recently
in individual SWNTs \cite{Morpurgo,KasumovPRB03,Herrero,Jorgenssen}.
The magnitude of the observed supercurrent has varied substantially.
In Refs. \onlinecite{KasumovPRB03,Herrero}, respectively, a
supercurrent on the order of 10x larger and 10x smaller than
theoretically expected was observed. Morpurgo \emph{et al.} and
J{\o}rgenssen \emph{et al.}, on the other hand, did observe only
increased conductance near zero bias.

In multiwalled carbon nanotubes (MWNT), supercurrents have been even
harder to achieve, presumably due to problems with disorder and
impurities. Enhanced conductance was observed by Buitelaar \emph{et
al.} \cite{BuitelaarPRL03} near zero bias, which was interpreted in
terms of multiple Andreev reflections (MAR) in the presence of
inelastic processes \cite{Vecino}. Recently, proximity induced
supercurrent has been observed by Kasumov \emph{et al.}
\cite{KasumovPRB03} as well as by Haruyama and coworkers
\cite{HaruyamaI,HaruyamaII}, most notably in multi-shell-contacted
tubes grown within nanoporous alumina templates \cite{HaruyamaII}.
Here we report proximity-induced superconductivity that is achieved
in an individual, diffusive MWNT using bulk(side)-contacted samples
with Ti/Al contacts. We find that the supercurrent can be smoothly
controlled by gate-voltage, via tuning of the diffusion constant,
and a good agreement is obtained using analysis based on long,
diffusive SNS junctions supplemented with phase diffusion effects,
modeled in terms of the resistively and capacitively shunted
junction model (RCSJ).

\begin{figure}

    \includegraphics[width=7.5cm]{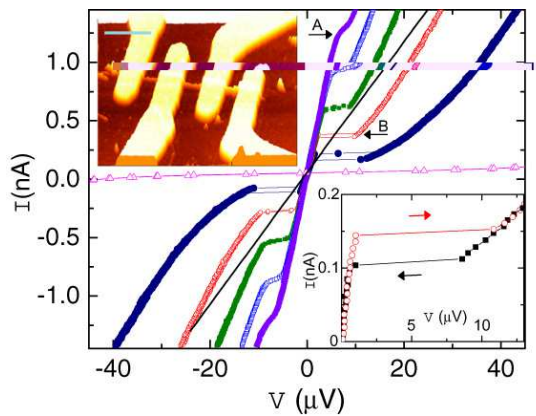}

    \caption{(color online) Current as a function of bias voltage
    $V$ at a few values of gate voltage
    $V_g=3.214$ V (\textcolor[rgb]{0.00,0.00,0.63}{$\bullet$});
    $V_g=3.220$ V (\textcolor[rgb]{1.00,0.00,0.00}{$\circ$});
    $V_g=3.226$ V (\textcolor[rgb]{0.00,0.50,0.00}{$\blacksquare$});
    $V_g=3.232$ V (\textcolor[rgb]{0.00,0.00,1.00}{$\square$});
    $V_g=3.244$ V (\textcolor[rgb]{0.44,0.00,0.87}{$\blacktriangle$});
    $V_g=3.344$ V (\textcolor[rgb]{1.00,0.00,0.50}{$\triangle$}).
    The solid straight line
    displays the normal state IV-curve measured in
    a magnetic field of $B=0.2$ T at $V_g=3.202$ V.
    The arrows A and B illustrate the determination of the maximum
    supercurrent $I_{cm}$ in the non-hysteretic and hysteretic cases.
    The inset on the lower right illustrates a
    magnification of the low voltage part of the IV-curve at
    $V_g=3.214$ V where clear hysteresis is visible and the
    critical current $I_{cm}=0.15$ nA.
    The inset on the upper left
    displays an AFM image of our sample (scale bar: 1 $\mu$m).
    The data were measured in a two-lead configuration on the
    middle section at $T=65$ mK.} \label{IV}

    \end{figure}

Our tube material, provided by the group of S. Iijima, was grown
using plasma enhanced growth without any metal catalyst
\cite{Koshio}. The tubes were dispersed in dichloroethane and, after
15 min of sonication, they were deposited onto thermally oxidized,
strongly doped Si wafers. A tube of 4 $\mu$m in length and 16.6 nm
in diameter was located with respect to alignment markers using a
FE-SEM Zeiss Supra 40. Subsequently, Ti contacts of width 550 nm
were made using standard overlay lithography: 10 nm titanium layer
in contact with the tube was covered by 70 nm Al in order to
facilitate proximity induced superconductivity in Ti at subkelvin
temperatures. Last, 5 nm of Ti was deposited to prevent the Al layer
from oxidation. The length of the tube section between the contacts
was 400 nm. The electrically conducting body of the silicon
substrate was employed as a back gate, separated from the sample by
150 nm of SiO$_2$. An AFM image of our sample is displayed in the
inset of Fig. \ref{IV}.

On our "dipstick" dilution refrigerator (Nanoway PDR50), the samples
were mounted inside a tight copper enclosure. The measurement leads
were filtered using an RC filter with time constant of 1 $\mu$s at
4.2 K, followed by twisted pairs with tight, grounded electrical
shields for filtering between still and the mixing chamber, while
the final section was provided by a 0.7-m-long Thermocoax cable on
the sample holder. In the measurements, current bias was employed
via a 100 M$\Omega$ room-temperature resistor when searching for
supercurrents, while voltage bias was employed when looking at
multiple Andreev reflections. In the latter case, also the
differential resistance $R=dV/dI$ was recorded using lock-in
techniques. The gap of the contact material, $\Delta_{lead}=139$
$\mu$eV, was found to be suppressed from our regular gap for
Aluminum leads ($\Delta_{reg}=200$ $\mu$eV) by 30 \%. Normal state
results were measured at $B=0.2$ T.

Our nanotube sample is slightly n-type doped initially, as in our
previous measurements on semiconducting MWNTs \cite{Fan06}. The
back gate capacitance $C_g=4$ aF was deduced from the measured
gate period of SET oscillations $\Delta V = 40$ mV. The total
capacitance of the 4 $\mu$m-long nanotube is estimated to be
$C_{\Sigma}=0.4$ fF which corresponds to a Coulomb energy of $E_C
\sim 2.5$ K. Since the Fermi-level of the tube shifts according to
$C_g/C_{\Sigma} e\Delta V_g $, the number of channels can be
varied only in the range of $N \simeq 0$ to a few over the
employed gate range of $V_g= -10 \ldots +10$ V. Most of the
observed changes in resistance $R(V_g)$ are then due to a
variation in the distribution of individual transmission
coefficients, not from the changes in $N$. There is a part of the
change, up to 50 \% at most (dependent on $V_g$), that can be
attributed to the Kondo effect. Other typical characteristics for
our samples are: the mean free path $l_{mfp} \sim 15$ nm and the
diffusion constant $D\sim v_F l_{mfp}=1 \cdot 10^{-2}$ m$^2$/s, as
deduced from the resistance. This means that a 400 nm long section
(see Fig. 1) is expected to be nearly in the long SNS junction
limit. We find several gate voltage values at which the normal
state resistance is in the range $R_n=15-20$ k$\Omega$. The
regions with good normal state conductance were found to vary by
$\pm20$ mV from day to day. This is attributed to variations in
the background charge. Occasionally, we also saw jumps of the
background charge which were on the order of $0.1 - 0.3$
electrons.

The measured IV-curves are illustrated in Fig. \ref{IV}. The gate
voltage has been varied over 100 mV which corresponds to a change of
2.5 electrons on the island. The shape is seen to change from a
state with linear IV-characteristics at small bias to a fully
"blockaded" one. The IV-curves in the first category display a
pronounced kink, followed by a plateau at voltages in the range of
$1-10$ $\mu$V; at intermediate gate voltages even hysteresis is
observed. This behavior is identified as proximity induced
superconductivity with the kink/plateau region indicating the
maximum measured supercurrent $I_{cm}$. A more detailed view of the
small bias regime is given in the lower right inset of Fig.
\ref{IV}. There is hysteresis at the intermediate values of the
critical currents. After a maximum hysteresis of $\Delta I = I_{cm}
- I_{retrap}= 54$ pA at $V_g=3.214$ V, the hysteresis is seen to
diminish as the current is lowered. The maximum product for $I_{cm}
R_n=22$ $\mu$eV $\ll \Delta_{lead}$.

According to the RCSJ model, there is hysteresis if the scaled
temperature $\Gamma=k_B T/E_J$ is small enough and the McCumber
parameter $\beta=(\omega_p R C_{tot})^2
>> 1$ where $\omega_p$ is the plasma frequency, $R$ is the shunt
resistance, and $C_{tot}$ is the total capacitance involved in the
plasma oscillation. To make an estimate, we take
$R=R_J=\frac{dV}{dI}\sim 2$ k$\Omega$ from the IV of the junction
above the plateau region, $C_{tot}=400$ fF, and $\omega_p=
\sqrt{\frac{2 e I_{c0}}{\hbar C_{tot}}}$, we get an estimate $\beta
\sim 24$ at $I_{c0}=5$ nA where $I_{c0}=\frac{2e E_J}{\hbar}$ is
taken at $E_J/k_B=120$ mK. This estimate is close to critical
damping and no hysteresis at $\Gamma \sim 1$ is to be expected
\cite{MK}. Nevertheless, a suppression of $I_{c0}= 5$ nA down to
$I_{cm}= 1.3$ nA takes place by thermal fluctuations. When $V_g$ is
tuned, $R_J$ increases more strongly than $\Gamma$, leading to the
appearance of hysteresis in the IV-curves due to a larger value for
$\beta$. Eventually the increase of $\Gamma$ with lowering $E_J$
takes over and the hysteresis disappears with decreasing $I_{cm}$ as
observed in Fig. \ref{IV}. Thus, RCSJ model can qualitatively
explain the main characteristics observed in Fig. \ref{IV}.

   \begin{figure}

    \includegraphics[width=7cm]{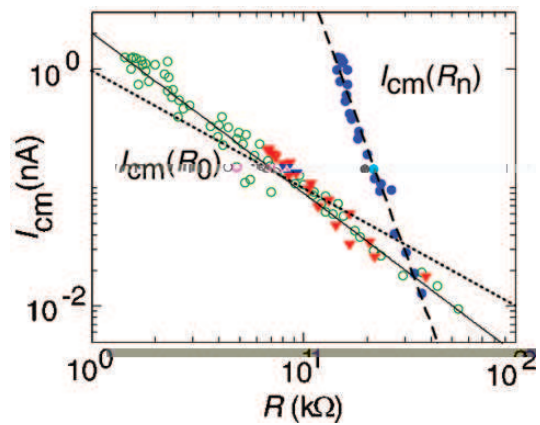}

    \caption{(color online) Maximum supercurrent $I_{cm}$ \textit{vs}. zero
    bias resistance $R_0$ and normal state resistance $R_n$ measured at $T=80$ mK.
    The filled red triangles and the open green circles displays
    $R_0$ at gate voltage $V_g = 5.98 \ldots 6.02$ V and $3.21 \ldots 3.33$ V, respectively.
    The filled blue circles denote $R_n$ at $V_g = 3.21 \ldots
    3.33$ V.
    The solid curve is a power law fit with $I_{cm}
\propto R_0^{-1.35}$ while the dashed line represents $I_{cm}
\propto R_n^{-4.93}$. The dotted curve displays the phase-diffusion
relation $I_{cm} \propto R_0^{-1}$ valid at $E_J << k_B
T$.}\label{IcvsR}

    \end{figure}

The supercurrent branch is found to display a finite resistance
which depends only weakly on bias (see the inset of Fig. \ref{IV}.
This weakness of bias dependence distinguishes our results from
those of Buitelaar et al \cite{BuitelaarPRL03} and of J{\o}rgenssen
et al \cite{Jorgenssen} who both observed a clear peak in the
conductance. The dependence of the measured supercurrent $I_{cm}$ on
 $R_0=\frac{dV}{dI}|_{V=0}$ as well
as on the normal state resistance $R_n$ is given in Fig.
\ref{IcvsR}. $I_{cm}$ could be tuned over two orders of magnitude
from 1.3 nA down to 10 pA, while the normal state resistance
increased only by a factor of two: from 15 to 35 k$\Omega$. We find
that the data can be fitted by a power law behavior $I_{cm} \propto
R_0^{-1.35}$.

The relatively large zero bias resistance, $R_0 > 1.4$ k$\Omega$, is
in agreement with the ordinary picture of phase diffusion
\cite{Tinkham,IZ}, which may coexist with hysteretic
IV-characteristics provided that the environment of the junction is
frequency dependent: at $\omega_p$, $Z_{env} \sim Z_0 = 377$
$\Omega$ stabilizing phase diffusion, while at low frequencies
$Z_{env} \gg Z_0$ \cite{MK}. Ingold \emph{et al.} \cite{IngoldPRB94}
have derived for the zero bias resistance due to phase diffusion
\begin{equation}\label{diffusion}
   R_0=\frac{Z_{env}}{I_0(E_J/k_B T)^2 - 1}
\end{equation}
where $I_0(x)$ represents a modified Bessel function. Subsequently,
Grabert \emph{et al.} have shown that Eq. (\ref{diffusion}) is
rather accurate, within a factor of $\sim 2$, even when the quantum
fluctuations are included \cite{GrabertEPL98}. In our analysis,
however, we stick to classical phase diffusion because the Coulomb
energy $E_c$ in our samples remains small owing to the large,
environmental shunting capacitance, which yields $k_B T/E_c \sim 50$
and negligible corrections from the charging effects.

In Ref. \onlinecite{IngoldPRB94}, the IV-characteristics was derived
in the limit $ E_J, eV << k_B T$, according to which there is a
simple relation between $I_{c0}$ and $I_{cm}$:
\begin{equation}\label{maxcurr}
   I_{cm}=\frac{E_J}{4 k_B T} I_{c0}.
\end{equation}
Thus, independent of $Z_{env}$, one expects $I_{cm} \propto E_J^2$
in the overdamped limit. From the maximum value of $I_{cm} = 1.3$
nA, we get $E_J/k_B=90$ mK at $T=65$ mK, which is at the limit of
applicability of Eq. (\ref{maxcurr}). Eq. (\ref{diffusion}) yields
$R_0 \propto E_J^{-2}$ in the limit $ E_J << k_B T$. Therefore,
using Eq. (\ref{maxcurr}), we get $I_{cm} \propto R_0^{-1}$, which
is seen to coincide quite well with our data in Fig. \ref{IcvsR}
where this dependence is shown as a dotted line.

The Josephson energy for a long diffusive junction, without
interaction effects, can be calculated from the equation
\cite{Zaikin81,DubosPRB}
\begin{equation}\label{Wilhelm}
I_{C0}=\frac{32}{3+2\sqrt{2}} \frac{\epsilon_{Th}}{eR_n}
\left(\frac{L}{L_T}\right)^3 \exp{\left(- \frac{L}{L_T} \right)}
\end{equation}

\noindent which is valid in the limit $\Delta/\epsilon_{Th}
\rightarrow \infty$ when $T \simeq 3 \epsilon_{Th}/k_B$ and where
$L_T=\sqrt{\hbar D/2 \pi k_B T}$. We are employing this formula in
our analysis as we are not aware of any appropriate theoretical
treatment of long, interacting SNS junctions. In SINIS structures,
perturbation analysis of the interacting case has shown that there
are logarithmic corrections that reduce Josephson coupling
\cite{Bruder94}, but this theory is not suitable for our case as the
contacts have a high transparency. We also neglect resonance effects
in our analysis, \emph{i.e.} the contribution that might be related
to Kondo effect.

\begin{figure}

    \includegraphics[width=8cm]{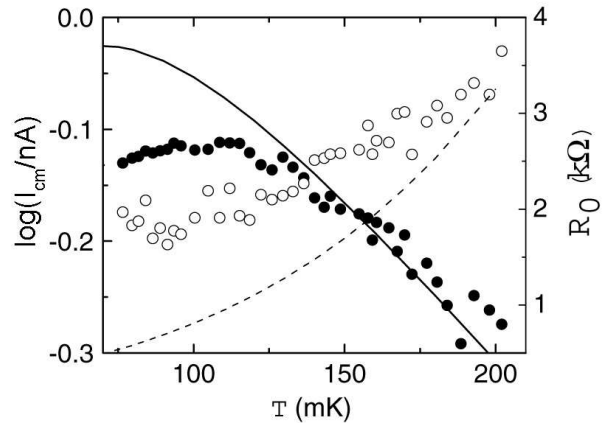}

    \caption{Temperature dependence of $I_{cm}$  ($\bullet$) and $R_0$ ($\circ$) measured
    at $V_g=3.489$ V. The solid curve is a fit obtained from
    Eqs. (\ref{maxcurr}) and (\ref{Wilhelm})
    using $D=2.2 \cdot 10^{-3}$ m$^2$/s, $L=0.4$ $ \mu$m, and
    $R_n= 17$ k$\Omega$. The dashed line displays
    $R_0$ calculated from Eq. (\ref{diffusion}) using $Z_{env}=400$ $\Omega$
    and $E_J(T)$ from the above $T$-dependence fit.
    } \label{Tdep}

    \end{figure}

By combining Eqs. (\ref{maxcurr}) and (\ref{Wilhelm}), we obtain an
analytical formula for $I_{cm}$ that has only one fitting parameter,
namely $D$ (or $L_T$ at certain $T$). In Fig. \ref{Tdep}, we compare
this formula with the measured temperature dependence of $I_{cm}$.
The solid curve in Fig. \ref{Tdep} is the result of the fit using
$L_T=\sqrt{80 \textrm{mK}/T}$ 188
 nm. This
 thermal length corresponds to $D=2.2 \cdot 10^{-3}$  m$^2$/s,
 yielding $\epsilon_{Th}=9.0$ $\mu$eV \cite{NOTE}.
 The
 discrepancy at the lowest temperatures may be an indication that
 Eq.  (\ref{maxcurr}) becomes invalid and numerical analysis based on
 the Ivanchenko-Zilberman theory should be done. Deviations at the
 lowest temperatures are also observed between the fit of Eq.
 (\ref{diffusion}) and the measured $R_0$ in Fig. \ref{Tdep}.

\begin{figure}

    \includegraphics[width=7.5cm]{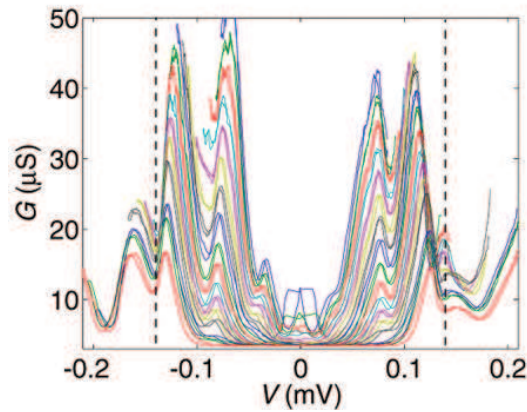}

    \caption{(color online) Conductance $G$ \emph{vs.} bias voltage $V$. Gate voltage $V_g$
    has been stepped over $V_g = 3.323 \ldots 3.355$ V in steps of 2
    mV. The dashed vertical lines indicate locations for the first Andreev
    reflection process if governed by unrenormalized  $\Delta_{lead}/e=139$
    $\mu$eV.
    } \label{MAR}

    \end{figure}

Fig. \ref{MAR} displays the differential conductance $G$ for the low
conductance IV-curves of Fig. \ref{IV} in the range $V_g=3.323
\ldots 3.355$ V. A sequence of maxima is seen, which are related to
multiple Andreev reflections (MAR) \cite{Octavio,Flensberg}. The MAR
peaks are more prominent than what we would expect for a long,
diffusive contact on the basis of a recent numerical analysis by
Cuevas et al. \cite{CuevasPRB06}. The dashed vertical lines indicate
locations for the first Andreev
    reflection process if it is governed by unrenormalized  $\Delta_{lead}/e=139$
    $\mu$V. Since Coulomb effects tend to shift MAR peaks upwards
    in voltage, the gap has to be modified at the interface  by
    20 \% downwards, roughly similar to findings by J{\o}rgenssen \textit{et
    al.} \cite{Jorgenssen} in SWNTs with Al/Ti contacts. Using
    $\tilde{\Delta}=0.8 \cdot \Delta_{lead}=111$ $\mu$eV we find that
    the main Andreev peaks are located at $\frac{2}{3}\tilde{\Delta}$
    and $\frac{2}{5}\tilde{\Delta}$, though the latter one does not
    coincide exactly to the expected location at $V>0$.

The weak gate dependence of the MAR lines is quite similar to that
found in Ref. \onlinecite{BuitelaarPRL03}. Between $\tilde{\Delta}$
and $2\tilde{\Delta}$ there is an additional peak that may be
connected to Thouless energy. As we approach the supercurrent region
by decreasing the gate voltage, $\epsilon_{Th}$ ($D$) increases and,
consequently, the peak should move towards $2\tilde{\Delta}$
\cite{CuevasPRB06}. In our experiment, however, the peak moves
towards $\tilde{\Delta}$. Notice also that in the data of Fig.
\ref{MAR}, the supercurrent peak near zero bias starts to develop
before any signs of higher order Andreev peaks. This seems to
contradict with the scenario of Vecino et al. \cite{Vecino} who
argue that inelastic processes enhance conductance due to higher
order MAR processes and lead to superconductor-like IV-curves with
hysteresis at small bias.

Since the number of transmission channels is rather small in our
sample, it is possible that the subgap transport is basically
dominated by one single channel. This might result, especially, from
the Kondo resonance that is known play a role in the conductance of
good quality MWNTs \cite{BuitelaarPRL03}. In fact, when comparing
the shape of the IV curves in Fig. \ref{IV} with the calculated
IV-curves for single-channel S-N-S contacts \cite{AverinBardasI}
rather good agreement is obtained for the range of transmissions
$\tau=0.3-0.7$. Therefore, even though a description using the model
of a long, diffusive junction seems to work well, presumably an
analysis based on a set of transmission channels would yield an even
better agreement.

In summary, we have observed proximity-induced superconductivity in
a diffusive PECVD-grown MWNT. The Josephson coupling energy could be
tuned by gate voltage via a change in the diffusion constant $D
\propto 1/R_n$. The model for long diffusive junctions was
successfully employed for calculating the dependence of $E_J$ on
$R_n$ and $T$. The measured IV curves (the zero bias resistance
$R_0$ and maximum supercurrent $I_{cm}$) could be understood using
analysis based on classical phase diffusion, which leads to a
decrease of $I_{cm}$ as $E_J^2$. At $T = 65$ mK, the largest
obtained Josephson coupling energy was 0.09 K.

We thank S. Iijima, A. Koshio, and M. Yudasaka for the carbon
nanotube material employed in our work. We wish to acknowledge
fruitful discussions with D. Gunnarsson, T. Heikkil\"a, F. Hekking,
P.-E. Lindel\"of, M.~Paalanen, B. Placais, C. Strunk, and A. Zaikin.
This work was supported by the TULE programme of the Academy of
Finland and by the EU contract FP6-IST-021285-2.


\begin{thebibliography}{99}
\bibitem{NOTE} The values for $D$ and $\epsilon_{Th}$ would be 10 - 20 \% higher if
one would employ a self-consistent numerical evaluation for the
product $I_{C0} R_n$. This has been neglected as the main error
comes from the lowest order estimate for $I_{cm}$.
\end{thebibliography}
\end{document}